\documentclass[sigplan,10pt]{acmart}
\renewcommand\footnotetextcopyrightpermission[1]{} 
\usepackage{multirow}

\usepackage[ruled,vlined]{algorithm2e}
\acmConference[PPoPP'21]{Principles and Practice of Parallel Programming}{Feb. 27th--Mar. 3rd, 2021}{Seoul, South Korea}
\acmYear{2021}
\acmISBN{} 
\acmDOI{} 
\startPage{1}

\setcopyright{none}

\usepackage{booktabs}   
\usepackage{subcaption} 

\begin{document}

\title[]{Fast and Efficient Parallel Breadth-First Search with Power-law Graph Transformation}         


\author{Zite JIANG}
\orcid{0000-0002-5680-5233}             
\affiliation{
  \institution{School of Computer Science and Technology, University of Chinese Academy of Sciences}            
}
\affiliation{
  \institution{SKL Computer Architecture, Institute of Computing Technology, Chinese~Academy~of~Sciences}            
}
\email{jiangzite19s@ict.ac.cn}          

\author{Tao LIU}
\affiliation{
  \institution{SKL Computer Architecture, Institute of Computing Technology, Chinese~Academy~of~Sciences}            
}
\email{liutao-ams@ict.ac.cn}         

\author{Shuai ZHANG}
\affiliation{
  \institution{School of Computer Science and Technology, University of Chinese Academy of Sciences}            
}
\affiliation{
  \institution{SKL Computer Architecture, Institute of Computing Technology, Chinese~Academy~of~Sciences}            
}
\email{zhangshuai-ams@ict.ac.cn}         

\author{Zhen GUAN}
\affiliation{
  \institution{SKL Computer Architecture, Institute of Computing Technology, Chinese~Academy~of~Sciences}            
}
\email{guanzhen@ict.ac.cn}         

\author{Mengting YUAN}
\affiliation{
  \institution{School of Computer Science, Wuhan University}            
}
\email{ymt@whu.edu.cn}         

\author{Haihang YOU}
\affiliation{
  \institution{SKL Computer Architecture, Institute of Computing Technology, Chinese~Academy~of~Sciences}            
}
\email{youhaihang@ict.ac.cn}         

\begin{abstract}
In the big data era, graph computing is widely used to exploit the hidden value in real-world graphs in various scenarios such as social networks, knowledge graphs, web searching, and recommendation systems. However, the random memory accesses result in inefficient use of cache and the irregular degree distribution leads to substantial load imbalance. Breadth-First Search (BFS) is frequently utilized as a kernel for many important and complex graph algorithms. In this paper, we describe a preprocessing approach using Reverse Cuthill-Mckee (RCM) algorithm to improve data locality and demonstrate how to achieve an efficient load balancing for BFS. Computations on RCM-reordered graph data are also accelerated with SIMD executions. We evaluate the performance of the graph preprocessing approach on Kronecker graphs of the Graph500 benchmark and real-world graphs. Our BFS implementation on RCM-reordered graph data achieves 326.48 MTEPS/W (mega TEPS per watt) on an ARMv8 system, ranking 2nd on the Green Graph500 list in June 2020 (the 1st rank uses GPU acceleration).
\end{abstract}



\keywords{Graph Computing, Breadth-First Search, Parallel Algorithms, SIMD}  

\maketitle

\section{Introduction}
Data-intensive graph-based computations play an important role in many modern big data graph applications such as social networks \cite{scott1988social}, knowledge graphs \cite{lin2015learning}, web searching \cite{page1999pagerank}, and recommendation systems \cite{pazzani2007content}. There is an increasing need for accelerating graph processing on modern parallel systems. The solutions to these graph applications typically involve classical graph algorithms such as shortest paths \cite{gallo1988shortest},connected components \cite{hirschberg1979computing}, spanning trees \cite{graham1985history} and centrality \cite{freeman1978centrality}. Breadth-First Search (BFS), an underlying graph algorithm, is frequently utilized as a kernel for these important and more complex graph algorithms \cite{ueno2016extreme}. Thus, accelerating BFS on modern parallel systems is crucial to the development of graph applications.

Most real-world graphs are large-scale but unstructured and sparse. One of the most notable characteristics of real-world graphs is the skewed \emph{power law} degree distribution \cite{gonzalez2012powergraph}: most vertices have a few neighbors while a few own a large number of neighbors. These characteristics present challenges for efficient parallel graph processing, such as load imbalance, poor locality, and redundant computations.

Apart from modifying the graph programming abstraction or changing the execution models on different architectures, reducing the irregularity of graph data also improves the performance of graph processing \cite{nodehi2018tigr}. For example, it's well-known that BFS has a bad temporal locality, but it's possible to transform irregular graphs to more regular ones to improve spatial locality and gain more performance.

We employ the Reverse Cuthill-Mckee (RCM) algorithm \cite{liu1976comparative,chan1980linear,azad2017reverse} to relabel vertices of a graph. The bandwidth of an adjacency matrix representing the graph is reduced, which implies improvements in the spatial locality of the graph data. Also, isolated vertices are excluded to reduce memory consumption. RCM-reordering makes logically related vertices clustered more close physically \cite{karantasis2014parallelization}. However, it introduces more serious load imbalance issues for BFS. We resolve the load imbalance strategies on the analysis of the features of the RCM-reordered graph data. To fully exploit the hardware resources and gain more performance improvements, we employ SIMD optimizations with NEON \cite{NEON:2020,seo2016efficient,camara2013fast} intrinsics on ARM systems.

The key contributions of this paper are:

1) We apply RCM-reordering on graph data to relabel vertex ID. The topology of the graph is unchanged. The isolated vertices are excluded for reducing memory consumption. The regularity of the graph is improved, thus reducing cache misses;

2) We present elaborate dynamic load balancing strategies for BFS to solve the load imbalance problem introduced by RCM-reorderging;

3) We employ SIMD optimization to gain further performance improvements;

4) We evaluate the performance of BFS on RCM-reordered Kronecker graphs and real-world graphs.

This paper is organized as follows. Section II discusses related work. Section III demonstrates the implementation details and optimization techniques of our optimized hybrid-BFS algorithms. Section IV illustrates the experimental results and Section V presents concluding remarks. 

\section{ Related Work}
\subsection{Graph Instance and Parallel BFS Algorithms}

The graph500 \cite{Graph500} benchmark is proposed to rank supercomputers based on their performance of data-intensive applications. It defines a new rate called traversed edges per second (TEPS) as the performance metric. The graph generated by the graph500 benchmark is a scale-free, low-diameter and small-world graph called Kronecker graph, similar to real-world social network graphs like twitter \cite{twitter} and friendster\cite{yang2012defining}. BFS is the most important kernel in the Graph500 benchmark.

From level-synchronous BFS to hybrid-BFS, parallel algorithms for BFS have been studied for decades on multicore systems \cite{bader2006designing, agarwal2010scalable, berrendorf2014level,yasui2013numa,yasui2014fast} and distributed memory systems \cite{bulucc2011parallel,yoo2005scalable, bulucc2017distributed, beamer2012direction, beamer2013distributed}. In recent years, most of the novel implementations for parallel BFS follow the structure of hybrid BFS and adapt the algorithm to better fit the underlying parallel architecture.

Given a graph $G=(V,E)$ with $n$ vertices, $m$ edges and a distinguished source vertex $s$. BFS traverses graph $G$ to discover all reachable vertices from $s$ level by level and produces an implicit \emph{breadth-first spanning tree} with root $s$ by maintaining a \emph{predecessor map}. All vertices at level $k$ are visited before any vertices at level $k+1$. The BFS \emph{frontier} is defined as the set of vertices in the current level. The complexity of sequential level-synchronous BFS algorithm is $O(m+n)$.

\begin{algorithm}
    \caption{Level-synchronous parallel BFS algorithm}
    \LinesNumbered
    \KwIn{$G=(V,E)$: undirected graph. \qquad \qquad \qquad \qquad $s$: source vertex.}
    \KwOut{$\pi$: predecessor map}
    $\pi(:)=-1, \pi(s)=s$\\
    $\textit{visited}(:)=\textit{false}, \textit{visited}(s)=\textit{true}$\\
    $F^{cur}=\{s\}, F^{next}=\emptyset$\\
    \While {$F^{cur} \neq \emptyset$}{
    \For {$v \in F^{cur}$ \textbf{in parallel}}{
    \For {$w \in \textit{neighbors}(v)$ }{
    \If {$\textit{visited}(w) \neq \textit{true} $ \textbf{atomic}}{
    $\textit{visited}(w)=\textit{true}$\\
    $\pi(w)=v$\\
    $F^{next}=F^{next} \cup \{w\} $\\
    }
    }
    }
    $F^{cur}=F^{next}$\\
    $F^{next}=\emptyset$
    }
\end{algorithm}

A multithread approach to \emph{level-synchronous} parallel BFS is demonstrated as Alg. 1. We assume that the input graph $G$ is both unweighted and undirected. The output $\pi$ is a predecessor map where $\pi(v)$ is the predecessor vertex on the shortest path from $s$ to $v$, or $-1$ if $v$ is not reachable from $s$. The \emph{visited} set maintains vertices already visited. $F^{cur}$ is the \emph{current frontier queue} initialized with $s$ and $F^{next}$ is the \emph{next frontier queue} initialized as an empty queue. At each iteration, each thread is assigned with a portion of vertices in $F^{cur}$, explores all neighbors of each vertex and finds unvisited neighbors and puts it at $F^{next}$ which will be traversed at the next level. Note that examinations on \emph{visited} and updates on both \emph{visited} and $F^{next}$ must be atomic to avoid race conditions and redundant computations. 

Real-world graphs such as social networks are usually low-diameter and scale-free. The total number of BFS steps is usually small, e.g. six to eight steps on a Kronecker graph. There are some extremely high degree vertices, causing certain frontiers to grow very fast. When the current frontier reaches its largest size, the majority of the graph has been visited and the following frontiers will shrink. There will be a great number of wasted checks on edges when traversing in the \emph{top-down} way (like Alg. 1) with a large frontier, which becomes the main bottleneck of level-synchronous parallel BFS.

\begin{algorithm}
    \caption{Hybrid parallel BFS algorithm}
    \LinesNumbered
    \KwIn{$G=(V,E)$: undirected graph. \qquad \qquad \qquad \qquad $s$: source vertex.}
    \KwOut{$\pi$: predecessor map}
    $\pi(:)=-1, \pi(s)=s$\\
    $\textit{visited}(:)=\textit{false}, \textit{visited}(s)=\textit{true}$\\
    $F^{cur}=\{s\}, F^{next}=\emptyset$\\
    $\textit{traversal\_policy}=\textit{TOP\_DOWN}$\\
    \While {$F^{cur} \neq \emptyset$}{
    \eIf {$\textit{traversal\_policy}=\textit{TOP\_DOWN}$}{
    \For {$v \in F^{cur}$ \textbf{in parallel}}{
    \For {$w \in \textit{neighbors}(v)$ }{
    \If {$\textit{visited}(w) \neq \textit{true} $ \textbf{atomic}}{
    $\textit{visited}(w)=\textit{true}$\\
    $\pi(w)=v$\\
    $F^{next}=F^{next} \cup \{w\} $\\
    }
    }
    }
    }{
    \For {$v \in V$ \textbf{in parallel}}{
    \If {$\textit{visited}(v)=\textit{false}$}{
    \For {$w \in \textit{neighbors}(v)$ }{
    \If {$w \in F^{cur}$}{
    $\textit{visited}(v)=\textit{true}$\\
    $\pi(v)=w$\\
    $F^{next}=F^{next} \cup \{v\} $\\
    \textbf{break}\\
    }
    }
    }
    }
    }
    $\textit{traversal\_policy}=\textit{update\_traversal\_policy}()$\\
    $F^{cur}=F^{next}$\\
    $F^{next}=\emptyset$\\
    }
\end{algorithm}

Beamer et al. \cite{beamer2012direction} proposed the hybrid BFS algorithm (Alg. 2) which combines the conventional top-down (level-synchronous) approach with a novel bottom-up approach. The hybrid algorithm employs the top-down approach on small frontiers and the bottom-up approach on large frontiers. When the frontier is large, the bottom-up approach will examine every edge attached to an unvisited vertex until finding a visited vertex in the frontier as its parent. Once a vertex has found a parent, it stops checking the rest of its neighbors, thus reducing the total number of edges examined (Alg. 2, line 15-21). Beamer et al. also explained a heuristic rule to determine when to switch between the top-down approach and the bottom-up approach by tuning two parameters: $\alpha$ and $\beta$.

\subsection{Graph Preprocessing and Transformation}
Many distributed graph processing systems such as Pregel\cite{malewicz2010pregel}, GraphLab\cite{low2012distributed}, and PowerGraph\cite{gonzalez2012powergraph}, have been proposed to handle graphs of extremely large scale by exploiting computation resources of clusters. However, there are still many challenges such as load imbalance, memory consumption, and synchronization overhead. Many techniques for graph preprocessing and transformation have been proposed to improve the efficiency of graph processing in recent years.

GraphChi\cite{kyrola2012graphchi}, a disk-based system, achieves efficient out-of-memory graph processing by well-designed partitioning and an asynchronous computation model which requires only a very small number of non-sequential accesses to the disk. GridGraph\cite{zhu2015gridgraph} improves the locality and reduces I/O operations with a novel grid representation for graphs and a fast streaming-apply graph processing model. GraphM\cite{zhao2019graphm} is an efficient storage system to amortize the storage consumption and the data access overhead between concurrent graph processing jobs. These graph processing systems focus on well-designed graph partitioning, fine-grained memory management, and efficient scheduling policy.

Tigr\cite{nodehi2018tigr} is a graph transformation framework that can effectively reduce the irregularity of real-world graphs to make these graphs better suited to GPU's SIMD execution. But Tigr changes the topology of a graph, which may affect the convergence of graph processing and alter the final results. Hao et al.\cite{wei2016speedup} proposed a graph ordering algorithm Gorder to keep frequently accessed nodes together locally for minimizing the cache miss ratio. However, Gorder is much slower than RCM\cite{liu1976comparative} and results in comparable performance for the BFS algorithm.

\subsection{Reverse Cuthill-Mckee Algorithm}
Many sparse matrix computations can be accelerated by reordering the matrix to reduce its bandwidth. Similarly, reordering vertices of a graph is crucial to minimize data size, maximize data locality and improve the performance of graph algorithms \cite{karantasis2014parallelization}. Several heuristic reordering algorithms are used in practice since computing the optimal reordering with a minimal bandwidth is NP-complete \cite{papadimitriou1976np}, such as Cuthill-Mckee (CM) \cite{cuthill1969reducing}, Reverse Cuthill-Mckee (RCM) \cite{liu1976comparative} and Slogan \cite{sloan1986algorithm}.

RCM reordering, a variant of CM reordering, is widely used to reduce the \emph{bandwidth} of a sparse symmetric matrix \emph{A}. The \emph{bandwidth} \cite{azad2017reverse} of matrix \emph{A} is defined as below:

\begin{equation}
    BW(\textbf{A}) = max\left\{i-j \Big| A_{ij} \ne 0, i > j \right\}
\end{equation}

Obtaining a reordering of rows or columns of \textbf{A} is equivalent to the process of relabeling vertices of graph $G(\textbf{A})$ associated with \textbf{A}. The RCM-reordered matrix usually has a smaller \emph{bandwidth}, where the non-isolated vertices are clustered closer and data locality is improved.

We will illustrate the decrease of cache misses while performing hybrid parallel BFS on RCM-reordered Kronecker graphs in section \textbf{4}.

\subsection{ARM Neon Technology}

ARM Neon technology \cite{NEON:2020}, the ARM Advanced SIMD (Single Instruction Multiple Data) architecture extension, provides thirty-two 128-bit vector registers on an ARMv8 system. Each register is capable of containing multiple lanes of data. SIMD instructions are utilized to perform the same operations in parallel on those multiple lanes of data. Many data-intensive applications can benefit from Neon technology, such as multimedia and signal processing, 3D graphics, speech, image processing, where fixed and floating-point performance is critical.

There are several ways to use Neon technology for programmers, including Neon intrinsics, Neon-enabled libraries, auto-vectorization by the compiler and hand-coded Neon assembler. The Neon intrinsics are a set of C and C++ functions supported by the Arm compiler and GCC. These intrinsics give programmers direct access to Neon instructions and offer substantial performance improvement without the need for hand-written assembly code. In this paper, we employ Neon intrinsics to improve the performance of parallel BFS.

\section{Implementation Details}
\subsection{Graph Representation}
The CSR (Compressed Sparse Row) representation is space-efficient and extremely fast on a single-node system, which provides constant-time access to neighbors of a vertex \cite{bulucc2017distributed}. When the graph fits comfortably into memory, it is recommended to use CSR for fast computations. In this paper, we use CSR to store the adjacency matrix representing the graph. There are mainly two arrays in this data structure: \emph{dst} holds neighbors' vertex ID of vertices in the graph; \emph{row\_starts} records the offset index of all neighbors of each vertex in the \emph{dst} array. Given a vertex $v_{0}$, the set of neighbors of $v_{0}$ in $dst$ are:

\begin{equation}
    \label{E1}
    \begin{aligned}
         & \textit{neighbors}(v_{0})=                                                                             \\
         & \left\{\textit{dst}[i] \Big| \textit{row\_starts}[v_{0}] \le i < \textit{row\_starts}[v_{0}+1]\right\}
    \end{aligned}
\end{equation}

Like the Kronecker graph in the Graph500 benchmark, we assume that the graph is undirected. So each edge is stored twice in both directions. We use 32-bit integers to represent the vertex ID, which is sufficient when the graph scale is at most 30 so that our multicore system with 384 GB memory can handle the Kronecker graph with scale 30.

\begin{algorithm}
    \caption{Reverse Cuthill-Mckee algorithm}
    \LinesNumbered
    \KwIn{$G=(V,E)$: undirected graph}
    \KwOut{$P$: permutation array}
    $V^{+} = $ all non-isolated vertices in $V$ sorted in increasing order of degree\\
    $\textit{sorted\_neighbors}(v) = $ neighbors of $v$ sorted in increasing order of degree\\
    $P(0) = V^{+}(0)$ \\
    $\textit{visited}(:)=\textit{false}$\\
    $\textit{min\_index} = 0$ \\
    $\textit{slow} = 0$, \textit{fast} = 0\\
    \While{$\textit{slow} < |V^{+}|$}{
    // find next unvisited vertex with the minimal degree\\
    \For{$i \in [\textit{min\_index},|V^{+}|)$}{
    \If{$\textit{visited}(V^{+}(i)) = \textit{false}$}{
            $P(\textit{slow}) = V^{+}(i)$\\
            $\textit{visited}(V^{+}(i)) = \textit{true}$\\
            $\textit{min\_index} = i + 1$\\
            $\textit{fast} = \textit{slow} + 1$\\
            \textbf{break}
            }
        }
        // explore next connected component and relabel \\
        // unvisited vertices in increasing order of degree \\
        \While{$\textit{slow} < \textit{fast}$}{
            \For {$v \in \textit{sorted\_neighbors}(P(\textit{slow}))$ }{
                    \If {$\textit{visited}(v) \neq \textit{true}$}{
                        $\textit{visited}(v)=\textit{true}$\\
                        $P(\textit{fast}) = v$\\
                        \textit{fast}++\\
                    }
                }
                \textit{slow}++\\
            }
        }
    $P=\textit{reverse}(P)$\\
    $P.\textit{append}(V \setminus V^{+})$\\
\end{algorithm}

\subsection{RCM Reordering}
A big problem with BFS is the bad locality because of the randomness of memory accesses on graph data. By reordering the vertex IDs to better fit in the access pattern of BFS, we can expect to achieve a higher locality. We apply RCM reordering to reduce the bandwidth of the adjacency matrix of the graph as a preprocessing step before running BFS. The RCM implementation is depicted in Alg. 3. 

The first labeled vertex strongly impacts the bandwidth of the permuted adjacency matrix. A frequently employed heuristic is to find a \textit{pseudo-peripheral vertex} with a high \textit{eccentricity} \cite{azad2017reverse}. However, to evaluate on a reproducible permutation array, we simply select the vertex with minimal degree as the fist labeled vertex (Alg. 3, line 11).

There are two prerequisite data for RCM: 
\begin{enumerate}
    \item $V^{+}$: All non-isolated vertices sorted in increasing order of degree. $V^{+}$ is used to find the vertex with minimal degree in the next connected component.
    \item $\textit{sorted\_neighbors}(v)$: Neighbors of $v$ sorted in increasing order of degree.
\end{enumerate}

RCM repeatedly explores all connected components in the graph, like BFS, level by level and relabels neighbors of the current vertex in increasing order of degree. In each connected component, the vertex with a minimal degree is first relabeled as \textit{slow} (Alg. 3 lines 9-15). \textit{slow} and \textit{fast} are both indices of the permutation array $P$. The RCM kernel repeatedly explores the neighbors of the vertex with an original ID of $P(\textit{slow})$, relabels unvisited neighbors and moves \textit{slow} and \textit{fast} forward until the current connected component is fully explored (breaking the while loop at Alg. 3 line 18). Then the consecutive connected components are explored in the same way. After relabeling all vertices in $V^{+}$, $P$ is reversed and all isolated vertices are appended to $P$. Then $P$ is the permutation array generated by RCM.

We recorded the number of connected components and the range of relabeled vertex ID for each connected component when the RCM-reordering is executed on the Kronecker graph with a scale from 21 to 30. Over 99.9\% of the non-isolated vertices are located in the last connected component and almost all other connected components are constituted by only two vertices connected to each other. The end ID of the last connected component is also the number of non-isolated vertices.

The original adjacency matrix, as shown in Fig. 1. (a), looks like the typical sparse symmetric matrix. According to the definition of bandwidth, It is estimated that its bandwidth is very close to $|V|$. After RCM reordering, the structure of the adjacency matrix becomes the shape of ``leaf'', which implies that the bandwidth is less than $|V^{+}|$. It can be considered that bandwidth is strong related to cache misses rate.

\begin{figure}[htbp]
    \centering
    \includegraphics[width=8cm]{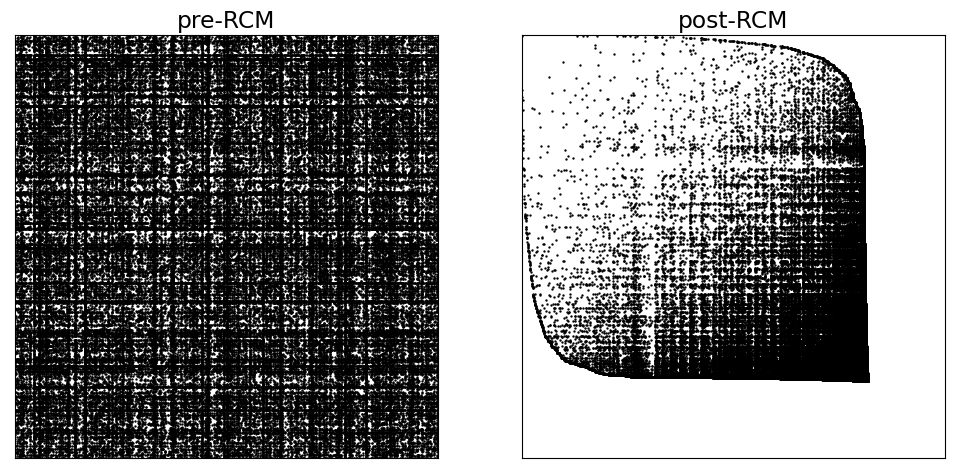}\\
    \caption{Structure comparison of adjacency matrices before and after RCM-reordering. The left side shows the adjacency matrix for the original Kronecker graph. The right side shows the adjacency matrix for the RCM-reordered graph.}
    \label{fig_1}
\end{figure}

\subsection{Top-Down Load Balancing}

Hybrid BFS on Kronecker graphs typically terminates after six to eight steps. These steps are divided into a \emph{growing} phase and a \emph{shrinking} phase according to the evolution of the \emph{frontier} in \cite{yasui2013numa}. Instead, We divide these steps into three phases and analyze the load balancing of each phase:

\emph{Phase 1:}
A \textit{top-down} phase in the first several steps. The frontier is small and usually contains high-degree vertices, which indicates the next frontier will be much larger. However, the degree distribution of frontier vertices is quite uneven, which will cause severe load imbalance among the threads. Some threads are idle, especially when the number of frontier vertices is less than the number of threads, which makes it impossible to make full use of all CPU cores. A solution to this problem is shown in Alg. 4. 

\begin{algorithm}
    \caption{Load-balanced top-down BFS}
    \LinesNumbered
    \KwIn{$G=(V,E) \text{: undirected graph}$. \quad $F^{cur} \text{: current frontier}$. \quad \quad \quad $t \text{: number of threads}$. $\textit{visited} \text{: the set of visited vertices}$. $\pi \text{: predecessor map}.$}
    \KwOut{$F^{next}\text{: next frontier}$.}
    $F^{next}=\emptyset$\\
    \For {all threads $T_i$ \textbf{in parallel}}{
    \For {$j \in [0, F^{cur})$ }{
    $v = F^{cur}[j]$\\
    $W = \textit{neighbors}_{i}(v, t)$\\
    \If {$i = j \mod t$}{
        $W = W \cup \textit{remaining\_neighbors}(v, t)$
    }
    \For {$w \in W$}{
    \If {$\textit{visited}(w) \neq \textit{true} $ \textbf{atomic}}{
    $\textit{visited}(w)=\textit{true}$\\
    $\pi(w)=v$\\
    $F^{next}=F^{next} \cup \{w\} $\\
    }
    }
    }
    }
\end{algorithm}

Like Alg. 1, Alg. 4 also uses a static scheduling policy. For each vertex in the current frontier, we divide its neighbors evenly into blocks of the same size and assign them to each thread. The remaining neighbors of the $j$th vertex are assigned to $(j \bmod t)$th thread in a Round-Robin way.

For a vertex $v_{0}$, the neighbors assigned to thread $i$ and the remaining neighbors are both obtained by calculating offset in \textit{dst} with thread ID in constant time as below:

\begin{equation}
    \label{E3}
    \begin{aligned}
         & \textit{neighbors}_{i}(v_{0}, t)=                                                                             \\
         & \left\{\textit{dst}[j] \Big| \textit{startpos}_{i} \le j < \textit{startpos}_{i} + \textit{workload} \right\} \\
         & \textit{remaining\_neighbors}(v_{0}, t) =                                                                        \\
         & \left\{\textit{dst}[j] \Big| \textit{startpos}_{t} \le j < \textit{row\_starts}[v_{0}+1] \right\}
    \end{aligned}
\end{equation}
where
\begin{equation*}
    \begin{aligned}
         & \textit{workload} = \frac{(\textit{row\_starts}[v_{0}+1] - \textit{row\_starts}[v_{0}])}{t} \\
         & \textit{startpos}_{k} = \textit{row\_starts}[v_{0}] + k \times \textit{workload}
    \end{aligned}
\end{equation*}

\begin{figure}[htbp]
    \centering
    \includegraphics[width=8cm]{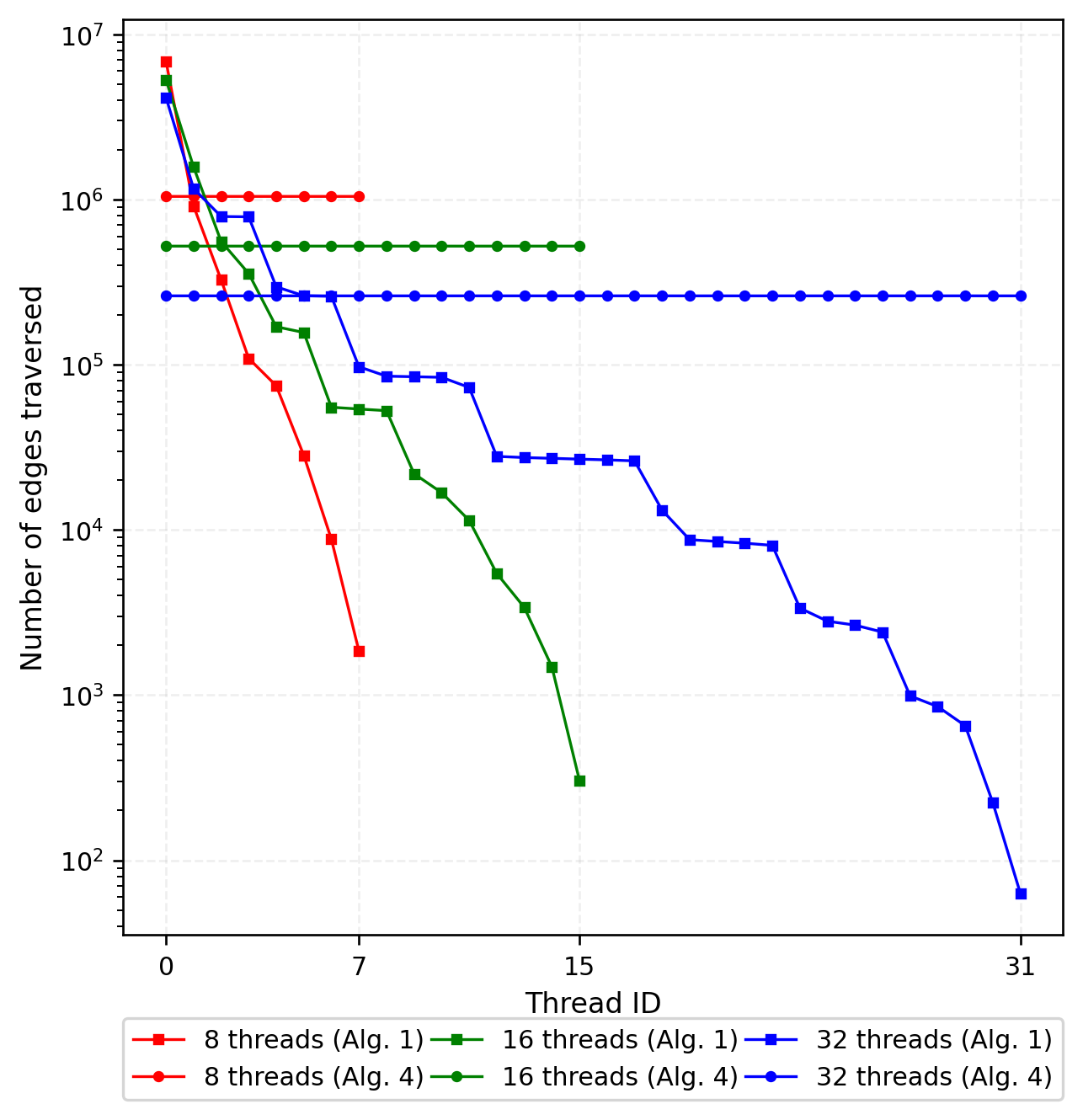}\\
    \caption{Comparison of distribution of edges among threads in a top-down step (scale 26, edgefactor 16, level 3). The edges are nearly evenly distributed to each thread for Alg. 4. However, for Alg. 1, most edges are assigned to the first several threads.}
    \label{fig_2}
\end{figure}
Fig. 2 shows the load balancing of \textit{top-down} kernel in Alg. 1 and Alg. 4. This experiment selected a Kronecker graph with scale 26 and edgefactor 16. We collect data on the distribution of the number of edges at level 3, a \textit{top-down} step, where 8364088 edges are traversed.

\emph{Phase 2:}
A \textit{bottom-up} phase in the middle several steps. The frontier is extremely larger than the ones in the other two phases. The benefits of RCM-reordering are best reflected in this phase, where most of the vertices are visited.

In the \textit{bottom-up} step, for each unvisited vertex, the \textit{bottom-up} kernel explores each of its neighbors until one neighbor in the frontier is found. The data structure of the frontier and next frontier is transformed into a bitmap so that we can check if a vertex is in the frontier and insert the vertex into the next frontier in constant time. The parameters \textit{visited}, $F^{cur}$ and $F^{next}$ are all bitmaps of length $|V|$. Each of these bitmaps is cache line aligned and divided into blocks of cache line size (64 bytes in this paper) without the risks of race conditions and \textit{false sharing}. And there is no need to use locks or atomic operations to keep thread safety.

Assuming there are $t$ threads, a simple static scheduling policy divides $V$ evenly into $t$ partitions and assigns one partition to each thread. This scheduling policy is efficient on random degree distribution before RCM-reordering but will cause load imbalance after RCM-reordering. We scrutinize the RCM-reordered graph data and it features an ascending degree distribution.

We use a descending partitioning policy to reduce differences in workload between partitions since the distribution of degrees of the RCM-reordered graph is ascending. One partition contains many blocks of cache line size. The sequence of the number of blocks in each partition is a descending arithmetic sequence. To achieve better load balancing, the number of generated partitions is much larger than the number of threads. At runtime, each thread continuously steals the next unprocessed partition and executes the \textit{bottom-up} kernel on that partition until all partitions have been processed.

We achieve good load balancing in the \textit{bottom-up} phase with the combination of static partitioning and dynamic work-stealing as demonstrated in Alg. 5. 
\begin{algorithm}
    \caption{Load-balanced bottom-up BFS}
    \LinesNumbered
    \KwIn{$G=(V,E) \text{: undirected graph}$. \quad $F^{cur} \text{: current frontier (bitmap)}$. \qquad $t \text{: number of threads}$. \qquad \qquad $\lambda \text{: partition factor.}$ $\textit{visited} \text{: the set of visited vertices (bitmap)}$. $\pi \text{: predecessor map}.$}
    \KwOut{$F^{next}\text{: next frontier (bitmap)}$.}
    $F^{next}=\emptyset$\\
    $S = \textit{get\_partitions}(G, \lambda, t)$\\
    $s = 0$\\
    \For {all threads $T_i$ \textbf{in parallel}}{
    $\textit{pos} = s\textit{.fetch\_add}(1)$\\
    \If {$\textit{pos} >= |S|$}{
        \textbf{break}
    }
    \For {$v \in S[\textit{pos}]$}{
    \If {$\textit{visited}(v)=\textit{false}$}{
    \For {$w \in \textit{neighbors}(v)$ }{
    \If {$w \in F^{cur}$}{
    $\textit{visited}(v)=\textit{true}$\\
    $\pi(v)=w$\\
    $F^{next}=F^{next} \cup \{v\} $\\
    \textbf{break}\\
    }
    }
    }
    }
    }
\end{algorithm}
A parameter $\lambda$ is introduced to determine the number of partitions:
\begin{equation}
    |S| = \lambda \times t
\end{equation}

$S[s]$ is the next partition to be processed. Each thread continuously steals a partition through a \textit{fetch\_add} operation on $s$ and executes the \textit{bottom-up} kernel until the value of $s$ is equal to or greater than $|S|$.

\emph{Phase 3:}
A \textit{top-down} phase in the last several steps. The frontier is small but relatively larger than in phase 1. Most of the vertices in the frontier are low-degree. It is efficient enough to employ the \textit{top-down} kernel in Alg. 1 in this phase.

\begin{algorithm}
    \caption{Load-balanced bottom-up BFS with workload reduction}
    \LinesNumbered
    \KwIn{$G=(V,E) \text{: undirected graph}$. \quad $F^{cur} \text{: current frontier (bitmap)}$. \qquad $t \text{: number of threads}$. \qquad \qquad $\lambda \text{: partition factor.}$ $\textit{visited} \text{: the set of visited vertices (bitmap)}$. $\pi \text{: predecessor map}.$}
    \KwOut{$F^{next}\text{: next frontier (bitmap)}$.}
    $\textit{neighbors}^{+}(v) = $ the highest-degree neighbor of $v$\\
    $\textit{neighbors}^{-}(v) = $ neighbors of $v$ sorted in decreasing order of degree (excluding $\textit{neighbors}^{+}(v)$)\\
    $F^{next}=\emptyset$\\
    $S = \textit{get\_partitions}(G, \lambda, t)$\\
    $s = 0$\\
    \For {all threads $T_i$ \textbf{in parallel}}{
    $\textit{pos} = s\textit{.fetch\_add}(1)$\\
    \If {$\textit{pos} >= |S|$}{
        \textbf{break}
    }
    $\textit{shrink}(S[\textit{pos}])$\\
    \For {$v \in S[\textit{pos}]$}{
    \If {$\textit{visited}(v)=\textit{false}$}{
    \For {$w \in \textit{neighbors}^{+}(v)$ }{
    \If {$w \in F^{cur}$}{
    $\textit{visited}(v)=\textit{true}$\\
    $\pi(v)=w$\\
    $F^{next}=F^{next} \cup \{v\} $\\
    }
    }
    }
    }
    $\textit{shrink}(S[\textit{pos}])$\\
    \For {$v \in S[\textit{pos}]$}{
    \If {$\textit{visited}(v)=\textit{false}$}{
    \For {$w \in \textit{neighbors}^{-}(v)$ }{
    \If {$w \in F^{cur}$}{
    $\textit{visited}(v)=\textit{true}$\\
    $\pi(v)=w$\\
    $F^{next}=F^{next} \cup \{v\} $\\
    \textbf{break}\\
    }
    }
    }
    }
    }
\end{algorithm}
\subsection{\textit{Bottom-up} workload reduction}
The major bottleneck of the hybrid BFS algorithm is the \textit{bottom-up} step. Reducing the workload is another way to accelerate the \textit{bottom-up} step apart from load balancing. 

We employ two methods to reduce the workload in a \textit{bottom-up} step in Alg. 6.

\emph{1) degree-aware BFS:}
The \textit{degree-aware} BFS\cite{yasui2014fast} is proposed to reduce the number of traversed edges in a \textit{bottom-up} step. It suggests that most traversed edges are concentrated in the first \textit{bottom-up} step and the number of traversed edges is affected by the ordering of the degree of each vertex's neighbors. And the descending ordering strategy is a better choice. As shown in Alg. 6, the neighbors of each vertex $v$ is sorted in descending degree and separated into the highest-degree neighbor $\textit{neighbors}^{+}(v)$ and the resting neighbors $\textit{neighbors}^{-}(v)$. The original loop of the \textit{bottom-up} kernel (Alg. 5, lines 8-15) is also separated into two loops (Alg. 6, lines 11-17 and lines 19-26). 

\emph{2) shrinking partitions:}
Once a thread steals a partition, it shrinks the partition before entering the \textit{bottom-up} kernel (Alg. 6 line 10,18). Each partition $[m, n]$ is shrunk to $[m', n']$ when all vertices in $[m, m')$ and $(n', n]$ are visited. The shrink operation does not introduce additional workload in the current step and helps reducing workload in the consecutive \textit{bottom-up} steps. Compared with the original graph representation, a partition is more likely to shrink after RCM-reordering because the visited vertices are spatially more closer to each other with the bandwidth reduction. 

\begin{figure*}[htbp]
    \centering
    \includegraphics[width=\linewidth]{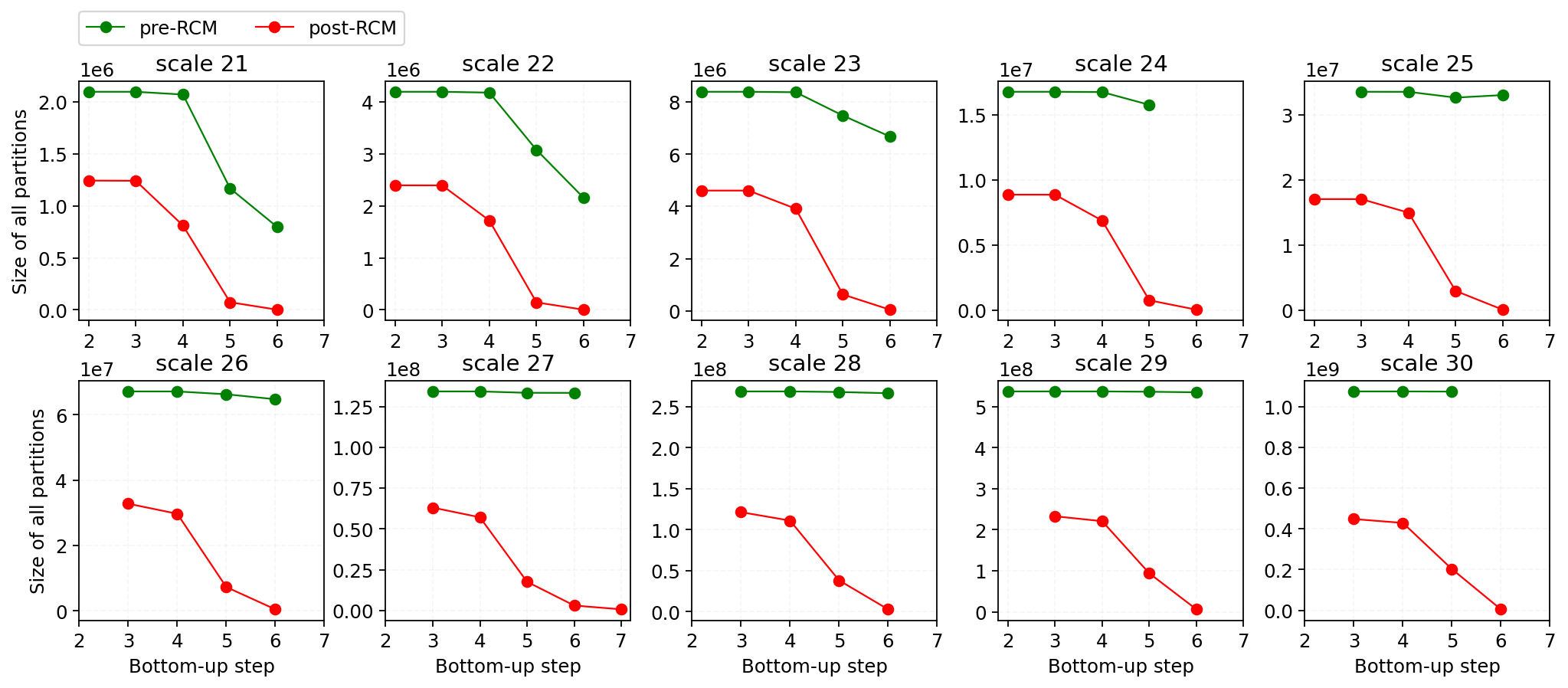}\\
    \caption{Size of all partitions at each bottom-up step before and after RCM-reordering. Each size is the average of values collected from the 64 rounds BFS (40 threads, edgefactor 16, $\lambda$ 20). The initial size of all partitions for the RCM-reordered graph is smaller than that of the original Kronecker graph because all isolated vertices are excluded. The former shows a significant reduction for all scales while the latter rarely reduces when the scale is greater than 24.}
    \label{fig_4}
\end{figure*}

Fig. \ref{fig_4} shows the size of all partitions at each \textit{bottom-up} step with a scale from 21 to 30. With 40 threads and $\lambda$ set to 20, the RCM-reordering enabled hybrid BFS demonstrates significant and stable reduction on partitions, especially from the 4th step to the 5th step. However, the RCM-reordering disabled version hardly reduces partitions with a scale greater than 23. In addition, the size of all partitions at the first bottom-up step of the RCM-reordering enabled version is much smaller than that of the RCM-reordering disabled version because all isolated vertices are already excluded.

The \textit{bottom-up} workload is reduced with the above two strategies.

\subsection{SIMD Optimizations}
There are almost no floating-point arithmetic operations in the BFS algorithm. Most of the operations are simple integer operations and bit manipulations. There are a large number of memory accesses, where read operations are much more than write operations. To make better use of the memory bandwidth, we locate two types of memory access intensive operations in our program and accelerate them with NEON intrinsics.

\emph{1) memory copy:}
Memory copy happens when (1) thread-local next frontier is copied to global frontier and (2) the \textit{bottom-up} next frontier is transformed into the \textit{top-down} frontier. We implemented a \textit{memcpy\_128()} function which encapsulates two 128-bit load and store NEON intrinsics  (\textit{vst1q\_u64()} and \textit{vld1q\_u64()}).

\emph{2) logical or} 
In the \textit{bottom-up} kernel, the newly visited vertex $v$ is updated into two bitmaps: \textit{visited} (Alg. 5 line 13) and $F^{next}$ (Alg. 5 line 15). We update the \textit{visited} outside the for loop at Alg. 5 line 9 through 128-bit logical or operations with the NEON intrinsic \textit{vorrq\_u64()}. But this method is only suitable for more regular graphs such that a preprocessing on graph data like RCM-reordering is required.

The NEON architecture provides full unaligned support for NEON data access. However, for the Cortex-A8 processor, specifying 128-bit or greater alignment saves one cycle per NEON instruction. We make most of the data 64 bytes aligned to take advantage of both cache efficiency and NEON acceleration.

\subsection{Parameter Tuning}
The parameters $\alpha$ and $\beta$ for hybrid BFS were tuned with the methodolgy as described in \cite{yasui2013numa}. We empirically set the ranges for $\alpha$ and $\beta$ to $[1,128]$ and $[1,32]$, then collect performance data of BFS. The optimal value of $(\alpha, \beta)$ is determined to be $(64, 8)$ by grid searching.

Another parameter $\lambda$ is tuned with 40 threads on Kronecker graphs with a scale from 21 to 30 and edgefactor 16. In general, the optimum value of $\lambda$ increases as the scale is larger. For graphs with scale less than 26, when $\lambda$ is greater than ten, the mean GTEPS decreases as $\lambda$ increases, because bigger $\lambda$ means more small partitions, which leads to more frequent work-stealing among threads. For graphs with a scale greater than 25, with $\lambda$ bigger than 20, the mean GTEPS fluctuates in a small range around 90 percent of the best performance for this scale. Thus, the value of $\lambda$ is recommended to be less than or close to 10 with a scale smaller than 26 and more than 20 with a scale bigger than 25.

\begin{table*}[!htb]
    \caption{cache miss comparison before and after RCM-reordering}
    \begin{center}

\resizebox{\textwidth}{!}{
        \begin{tabular}{|c|l|l|l|l|l|l|l|l|}
            \hline
            \multirow{2}{*}{\textbf{scale}} & \multicolumn{3}{c|}{\textbf{cache references }} & \multicolumn{3}{c|}{\textbf{cache misses }} & \multicolumn{2}{c|}{\textbf{cache miss rate (\%)}}                                                                                                                                                         \\
            \cline{2-9}
                                            & \textbf{pre-RCM (10\textsuperscript{9})}        & \textbf{post-RCM (10\textsuperscript{9})}   & \textbf{reduction (\%)}                            & \textbf{pre-RCM (10\textsuperscript{9})} & \textbf{post-RCM (10\textsuperscript{9})} & \textbf{reduction (\%)} & \textbf{pre-RCM} & \textbf{post-RCM} \\
            \hline
            21                              & 4.5783                                          & 1.6449                                      & 64.07                                              & 0.2111                                   & 0.0474                                    & 77.55                   & 4.59             & 2.88              \\
            22                              & 9.6919                                          & 3.2398                                      & 66.57                                              & 0.4654                                   & 0.1127                                    & 75.78                   & 4.80             & 3.47              \\
            23                              & 18.9666                                         & 6.2313                                      & 67.15                                              & 0.8536                                   & 0.2795                                    & 67.26                   & 4.49             & 4.48              \\
            24                              & 38.3984                                         & 11.8044                                     & 69.26                                              & 1.6832                                   & 0.5653                                    & 66.42                   & 4.38             & 4.79              \\
            25                              & 80.5952                                         & 23.4695                                     & 70.88                                              & 3.6657                                   & 1.4753                                    & 59.75                   & 4.55             & 6.29              \\
            26                              & 161.2918                                        & 47.5949                                     & 70.49                                              & 6.8352                                   & 3.7127                                    & 45.68                   & 4.23             & 7.80              \\
            27                              & 331.3712                                        & 89.7782                                     & 72.91                                              & 15.6341                                  & 6.6334                                    & 57.57                   & 4.71             & 7.39              \\
            28                              & 648.0242                                        & 165.1345                                    & 74.52                                              & 28.5306                                  & 11.5033                                   & 59.68                   & 4.40             & 6.96              \\
            29                              & 1344.9668                                       & 348.2868                                    & 74.10                                              & 55.2038                                  & 32.1911                                   & 41.69                   & 4.10             & 9.23              \\
            30                              & 2600.5620                                       & 676.3802                                    & 73.99                                              & 111.8545                                 & 68.3663                                   & 38.88                   & 4.30             & 10.09             \\
            \hline
        \end{tabular}
		}
        \label{T4}
    \end{center}
\end{table*}

\section{PERFORMANCE EVALUATION}
\subsection{Evaluation Platform}
We evaluate the performance of our algorithm on an ARMv8 system. The experimental environments are listed in Table \ref{platform}.

\begin{table}[htbp]
    \caption{Evaluation Platform}
    \begin{center}
        \begin{tabular}{ll}
            \hline
            \textbf{Processor}       & Qualcomm Centriq 2434, 2.3 GHz \\
            \textbf{Socket}          & 1                              \\
            \textbf{Cores}           & 40                             \\
            \textbf{Threads}         & 40                             \\
            \textbf{Cache(L1/L2/L3)} & 3.75 MB / 10 MB / 50 MB         \\
            \textbf{SIMD}            & 128 bits                       \\
            \textbf{Mem.}            & 384 GB (DDR4 2666 MHz)         \\
            \textbf{TDP}             & 110 W                          \\
            \textbf{OS, Compiler}    & Centos 7, gcc-7.3.1            \\
            \hline
        \end{tabular}
        \label{platform}
    \end{center}
\end{table}

\subsection{Graph Instances}
The performance of our implementation is measured on Kronecker graphs generated by the official Graph500 benchmark. Initiator parameters for the Kronecker graph generator $(A,B,C,D)$ are set to $(0.57, 0.19, 0.19, 0.05)$. The graph size is determined by two parameters: \textit{scale} and \textit{edgefactor}. The total number of vertices is $N=2^{\textit{scale}}$ and the number of edges is $M=N \times \textit{edgefactor}$. There are some self-loops, duplicated edges and isolated vertices in the graph.

\subsection{Performance Overview}
\begin{figure*}[htbp]
    \centering
    \includegraphics[width=\linewidth]{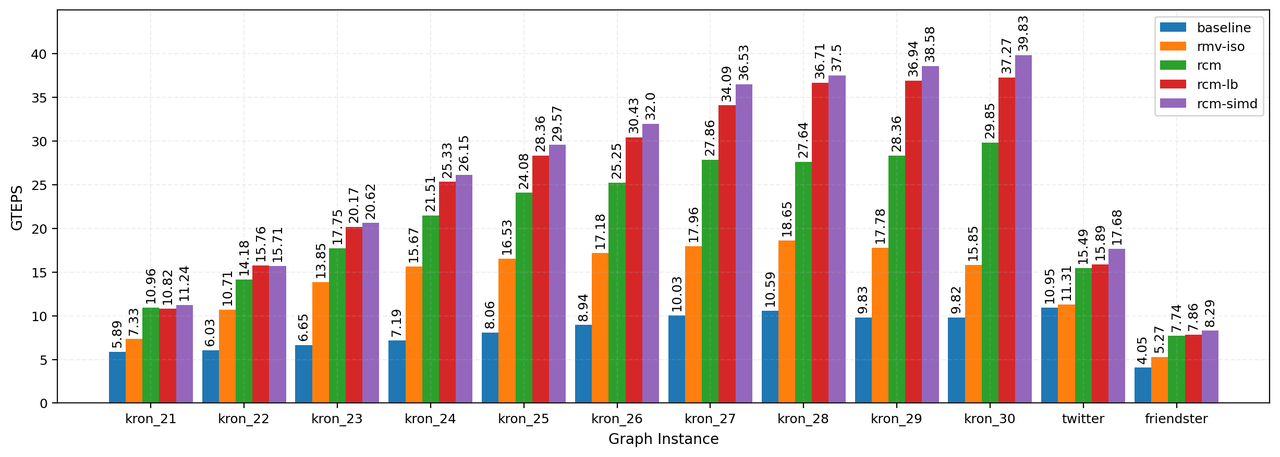}\\
    \caption{Ablation experiments on different optimizations. \emph{1. baseline: hybrid bfs; 2.rmv-iso: baseline + remove isolated vertices; 3.RCM: rmv-iso +RCM; 4.rcm-lb: RCM+load balance; 5.rcm-simd: rcm-lb + simd}.}
    \label{overview}
\end{figure*}

Ablation experiments are tested for each optimization strategy as shown in Fig. \ref{overview}. In fact, every optimization can improve the performance independently and works well in kronecker graphs, twitter \cite{yang2012defining} and friendster \cite{twitter}.

\subsection{Cache Miss Reduction}
Table 3 shows the cache misses reduction after applying RCM-reordering on original Kronecker graphs. Our hybrid BFS algorithm runs on Kronecker graphs with a scale from 21 to 30 and edgefactor 16 using a single thread. We collect statistics about cache references and cache misses during running BFS on 64 randomly selected source vertices (after generating Kronecker graph and RCM-reordering) and calculate the corresponding cache miss rate.

Both cache references and cache misses are decreased a lot with RCM-reordering enabled. The cache references decrease at an ascending speed and the cache misses decrease at a descending speed. However, when RCM-reordering is enabled, the cache miss rate becomes high as scale increases and exceeds that when RCM-reordering is disabled with a scale greater than 24. The reason is that the percentage of non-isolated vertices decreases as the scale increases and cache references are greatly reduced. Nevertheless, the performance is improved since the absolute amount of cache references and cache misses instead of the cache miss rate account for the running time.

\subsection{Strong Scaling}
Strong scaling for original hybrid BFS and RCM-reordering enabled hybrid BFS on a Kronecker graph with scale 30 and edgefactor 16 are illustrated in Table 4 and Fig. 6. Compared with original hybrid BFS, RCM-reordering enabled hybrid BFS demonstrates a speedup over 4 times and better scalability.

\begin{table}[!htb]
    \caption{Strong Scaling comparison (scale 30, edgefactor 16)}
    \begin{center}
        \begin{tabular}{|c|c|c|c|c|}
            \hline
            \multirow{2}{*}{\textbf{Threads}} & \multicolumn{2}{c|}{\textbf{Hybrid BFS}} & \multicolumn{2}{c|}{\textbf{RCM + Hybrid BFS}}   \\
            \cline{2-5}
                                              & \textbf{GTEPS}                           & \textbf{Speedup}                               & \textbf{GTEPS} & \textbf{Speedup} \\
            \hline
            1    & 0.235   & 1.0     & 0.965              & 1.0                \\
            2    & 0.414   & 1.8     & 1.963              & 2.0                \\
            4    & 0.846   & 3.6     & 4.011              & 4.2                \\
            8    & 1.626   & 6.9     & 7.821              & 8.1                \\
            16   & 3.256   & 13.9     & 15.284              & 15.8                \\
            32   & 6.068   & 25.8     & 29.115              & 30.2                \\
            40   & 7.283   & 31.0     & 39.834              & 41.3                \\
            \hline
        \end{tabular}
        \label{T5}
    \end{center}
\end{table}

The strong scaling of the RCM-reordering enabled hybrid BFS with 40 threads is approximately 41 times faster than sequential computation, achieving a superlinear scaling. The reason is that the number of edges precisely traversed is not as large as the total number of edges generated by the Kronecker generator with plenty of duplicated edges. 



\begin{figure*}[!h]
    \centering
    \includegraphics[width=\linewidth]{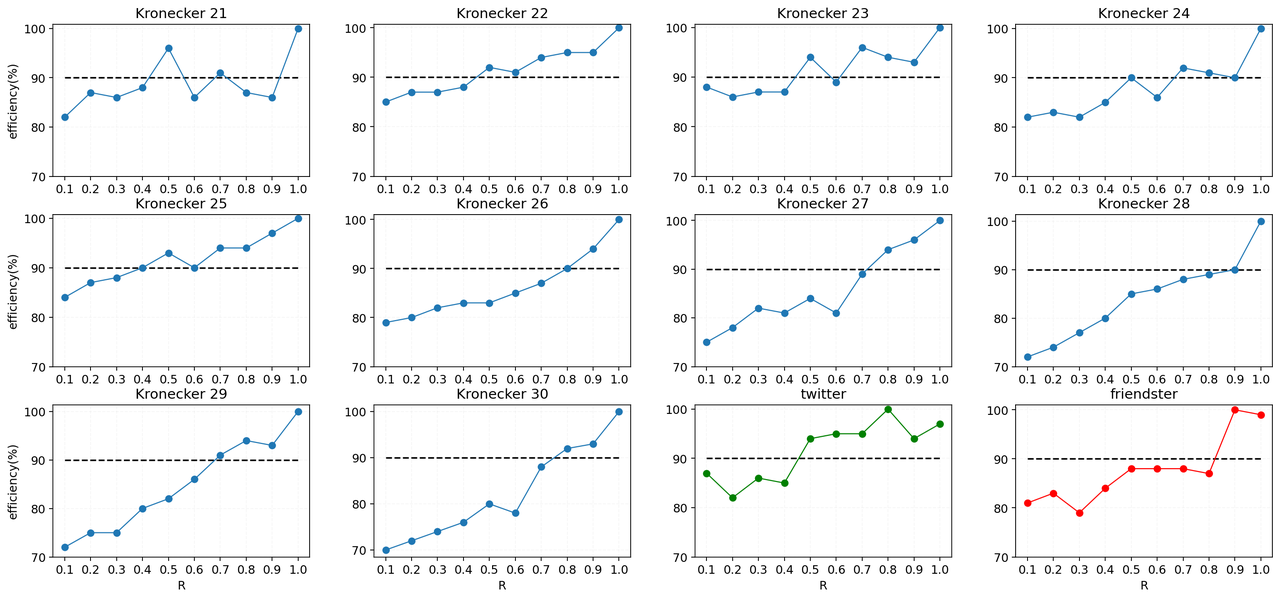}\\
    \caption{Partial RCM on Kronecker, twitter and friendster graphs.}
    \label{partial_rcm}
\end{figure*}

\subsection{Partial RCM}
Admittedly, a RCM-based algorithm will increase the overall algorithm time, and the overhead of RCM is much greater than BFS time in Table \ref{RCM_time}.

\begin{table}[htbp]
    \caption{RCM time cost in BFS}
    \begin{center}
\scalebox{0.9}{
        \begin{tabular}{|c|c|c|c|}
            \hline
		\textbf{graph} & \textbf{RCM time(s)} & \textbf{RCM BFS time(s)}  & \textbf{BFS time(s)} \\ \hline
            $kronecker$         &  836.707    & 0.425441             & 1.08366              \\  \hline
            $twitter$          &  35.2149    & 0.0885213            & 0.134154              \\  \hline
            $friendster$       &  50.2221    & 0.226269             & 0.342534              \\ \hline
        \end{tabular}
        \label{RCM_time}
}
    \end{center}
\end{table}

Partial RCM is proposed to reduce RCM time cost while maintaining good memory access locality. A $p$-partial RCM is to only reorder the top-$p$ percents vertices rather than to reorder all vertices. Intuitively, as the ratio increases, efficiency will increase. And the reduction in time is linearly related to the reduction in proportion.

However, considering the influence of vertices proportion from 0.1 to 1 on the efficiency of RCM BFS, the results show that partial RCM have better performance with proper ratio value in twitter and friendster graphs in Fig. \ref{partial_rcm}. The efficiency of each ratio is obtained by averaging multiple experiments, and the 100\% efficiency corresponds to the fastest RCM BFS.

\section{CONCLUSION}

In this paper, inspired by research on the Reverse Cuthill-Mckee algorithm for reducing the bandwidth of a sparse matrix, we combine hybrid BFS with RCM-reordering to achieve better efficiency for BFS. The RCM-reordering demonstrates great advantages such as the reduction in memory accesses, improvements on data locality and significant workload reduction between consecutive \textit{bottom-up} steps. In addition, we employ SIMD optimizations to fully exploit the hardware resources. Our optimized BFS implementation achieves 3 times speedup over the approach without RCM-reordering and shows a near-linear strong scaling on an ARMv8 system with NEON support. The resulting performance of 39.83 GTEPS is able to rank 79th on the Graph500 list in June 2020. Also, with an average power of 122 watts, our implementation achieves a performance of 326.48 MTEPS/W and ranks 2nd on the Green Graph500 list in June 2020. The optimizations also can be extended to real-world graphs. 

With the performance improvements for the BFS algorithm, the results of this work demonstrate potential performance improvement for more complex traversing-based graph algorithms, such as shortest paths, connected components, and spanning trees. Therefore, this work is also valuable to accelerate many real-world graph applications and achieve higher throughput for graph processing.

\bibliography{main}

\newpage


\end{document}